\documentclass[prd,showpacs,preprintnumbers]{revtex4-1}
\usepackage{graphicx}
\usepackage{hyperref}
\usepackage{amsfonts}
\usepackage{amsmath,amssymb}

 \def\be{\begin{equation}}
 \def\ee{\end{equation}}
 \def\bes{\begin{eqnarray}}
 \def\ees{\end{eqnarray}}

 \def\E#1{\epsilon_#1}
 \def\A{\mathcal{A}}

 \def\N{\mathcal{N}}
 \def\O{\mathcal{O}}
 
 \def\2{\frac{1}{2}}
 \def\4{\frac{1}{4}}

\catcode`\@=11
\def\@citex[#1]#2{%
\if@filesw \immediate \write \@auxout {\string \citation {#2}}\fi
\@tempcntb\m@ne \let\@h@ld\relax \def\@citea{}%
\@cite{%
  \@for \@citeb:=#2\do {%
    \@ifundefined {b@\@citeb}%
      {\@h@ld\@citea\@tempcntb\m@ne{\bf ?}%
      \@warning {Citation `\@citeb ' on page \thepage \space
undefined}}%
      {\@tempcnta\@tempcntb \advance\@tempcnta\@ne%
      \@tempcntb\number\csname b@\@citeb \endcsname \relax%
      \ifnum\@tempcnta=\@tempcntb 
it
        \ifx\@h@ld\relax%
          \edef \@h@ld{\@citea\csname b@\@citeb\endcsname}%
        \else%
          \edef\@h@ld{\ifmmode{-}\else--\fi\csname
b@\@citeb\endcsname}%
        \fi%
      \else
        \@h@ld\@citea\csname b@\@citeb \endcsname%
        \let\@h@ld\relax%
      \fi}%
    \def\@citea{,\penalty\@highpenalty\,}%
  }\@h@ld
}{#1}}

\def\@citeb#1#2{{[#1]\if@tempswa , #2\fi}}
\def\@citeu#1#2{{$^{#1}$\if@tempswa , #2\fi }}
\def\@citep#1#2{{#1\if@tempswa , #2\fi}}

\begin{document}
\preprint{UTHET-11-1002}

\title{Gauge theory one-loop amplitudes and the BCFW recursion relations}

\author{Savan Kharel}
 \email{skharel@tennessee.edu}
\author{George Siopsis}
 \email{siopsis@tennessee.edu}
\affiliation{%
Department of Physics and Astronomy,
The University of Tennessee,
Knoxville, TN 37996 - 1200, USA}
\date{October 2011}%

\begin{abstract}
We calculate gauge theory one-loop amplitudes with the aid of the complex shift used in the
Britto-Cachazo-Feng-Witten (BCFW) recursion relations of tree amplitudes. We apply the shift to the integrand and show that the contribution from the limit of infinite shift vanishes after integrating over the loop momentum, with a judicious choice of basis for polarization vectors.
This enables us to write the one-loop amplitude in terms of on-shell tree and lower-point one-loop amplitudes. Some of the tree amplitudes are forward amplitudes. We show that their potential singularities do not contribute and the BCFW recursion relations can be applied in such a way as to avoid these singularities altogether. We calculate in detail $n$-point one-loop amplitudes for $n=2,3,4$, and outline the generalization of our method to $n>4$.
\end{abstract}


\maketitle

\section{Introduction}
There are several reasons to improve on our understanding of scattering amplitudes in gauge theories, ranging from the development of an efficient and accurate calculation of standard model processes that occur in high energy accelerators such as the Large Hadron Collider (LHC),
to
formal developments, such as understanding the properties of quantum field theory and
quantum gravity.

In the last few years, there has been extraordinary progress in the study of scattering amplitudes. We learned  that the scattering amplitudes of gravity and gauge theories have more structure and symmetries than are
manifest in the Lagrangian.
One of the first extraordinary properties of scattering amplitudes was discovered in the mid-eighties by Parke and
Taylor who found an extremely simple and compact expression for
Maximally-Helicity-Violating (MHV) amplitudes \cite{Parke:1986gb}. The modern renaissance in the study of scattering amplitudes was led by an
important conceptual development due to Witten \cite{Witten:2003nn} who observed that the structure of gauge theory scattering amplitudes
is very simple in twistor space. For recent reviews of scattering amplitudes, see, e.g., \cite{Peskin:2011in, Dixon:2011xs}.

Witten's seminal work inspired an important contribution by Cachazo, Svrcek, and Witten \cite{Cachazo:2004kj} and its extension, the Britto-Cachazo-Feng-Witten (BCFW) recursion relations \cite{Britto:2005fq}. In the BCFW prescription, a pair of the external momenta in a tree amplitude are analytically
continued into the complex plane, turning the amplitude into a meromorphic function. Thus, these amplitudes are shown to be determined
by the residues of their poles. The BCFW technique exploits this property in order to recursively construct physical amplitudes from irreducible three-point amplitudes.
However, in order to effectively use recursion relation, the residue of the pole at infinity must vanish. This is the case in gauge theories and gravity, but not in generic field theories \cite{ArkaniHamed:2008yf}. In the last few years much progress has been realized in our understanding of scattering amplitudes based on the BCFW recursion relation. For example, the BCFW recursion relations have been applied to amplitudes involving gravitons \cite{Bedford:2005yy, Cachazo:2005ca, Benincasa:2007qj, BjerrumBohr:2005jr}, string theory \cite{Cheung:2010vn, Boels:2010bv, Fotopoulos:2010cm}, and anti-de Sitter (AdS) space \cite{Raju:2010by}.

The extension of the BCFW recursion relations to loop amplitudes is not straightforward. Loop amplitudes receive, in general, a non-vanishing contribution from the pole at infinity. They also possess cuts, in addition to poles, which makes the application of Cauchy's theorem more cumbersome.
In the mid-nineties, powerful on-shell unitarity methods  were developed for the calculation of scattering amplitudes  \cite{Bern:1994zx, Bern:1994cg} (for a review, see  \cite{Bern:1996je}). A generalization of the BCFW recursion relations and the unitarity method to loop amplitudes was then considered  \cite{Bern:2005hs, Bern:2005cq, Berger:2008sj,Dunbar:2010wu}. An alternative approach, in which one applies the BCFW recursion relations to the {\em integrand} of the loop amplitude, was recently discussed \cite{Boels:2010nw}. In the case of $\mathcal{N} =4$ super Yang-Mills gauge theory, all loop amplitudes were thus obtained in the planar limit \cite{ArkaniHamed:2010kv}.

In this paper, we re-visit the application of BCFW recursion relations to the {\em integrand} of gauge-theory loop amplitudes. We concentrate on one-loop amplitudes, although our results can be generalized to higher loop order. We show that the contribution of the pole at infinite complex shift can be made to vanish, after integrating over the loop momentum, by a judicious choice of basis for the polarization vectors.
This enables us to express one-loop amplitudes in terms of tree amplitudes and lower-point one-loop amplitudes. The tree amplitudes include forward amplitudes which are plagued by divergences, in general. We show that these potential divergences do not contribute and discuss how the BCFW recursion relations can be applied so as to avoid the divergences, thus reducing the one-loop amplitudes to three-point tree amplitudes.

We perform the calculation in detail for two-point (section \ref{sec:2}), three-point (section \ref{sec:3}), and four-point (section \ref{sec:4}) one-loop amplitudes. In section \ref{sec:5}, we outline the generalization of our method to one-loop amplitudes with $n>4$. In section \ref{sec:6}, we summarize our conclusions. We work with color ordered amplitudes throughout, to simplify the discussion.

\section{Two-point loop amplitude}
\label{sec:2}

In this section, we consider a two-point one-loop amplitude. Ignoring group theory factors, it can be written as an integral over the loop momentum,
\be A_2^{\mathrm{1-loop}}(k_1, \E1 ; -k_1 , \E2 ) = \int \frac{d^{2\omega} l}{(4\pi)^{2\omega}} \mathcal{A}_2^{\mathrm{1-loop}}(k_1, \E1 ; -k_1 , \E2 )~, \ee
where $\omega$ is a dimensional regularization parameter, and the two polarization vectors are null, with $\E1\cdot k_1 = \E2\cdot k_1 = 0$.
The momentum $k_1$ is off shell.

To apply the BCFW recursion relations, we
shift $k_1 \mapsto k_1 + z \E1$.
Consequently, we shift the second polarization vector,
\be\label{eq2} \E2\mapsto \E2' \equiv \E2- z \frac{\E2\cdot \E1}{k_1^2} k_1~. \ee
We will use the background gauge \cite{Abbott:1983zw} in order to compute this amplitude.
\begin{figure}[ht]
\begin{center}
\includegraphics[width=9cm]{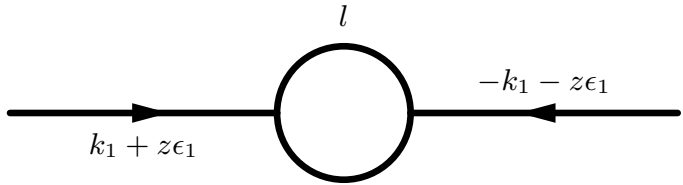}
\caption{Diagram contributing to the two-point one-loop amplitude.}
\label {TwoP}
\end{center}
\end{figure}
There is only one diagram that contributes to this amplitude (figure \ref{TwoP}).

\begin{figure}[ht]
\begin{center}
\includegraphics[width=15cm]{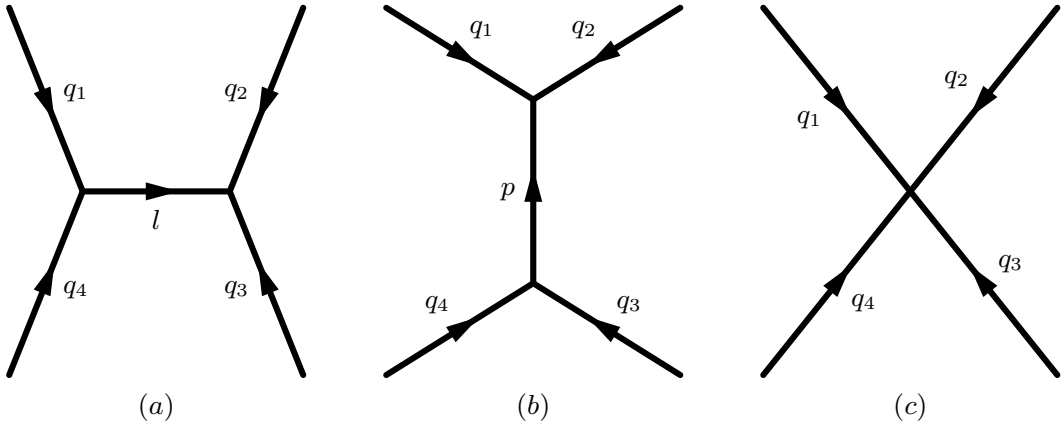}
\caption{Four-point tree diagrams contributing to a two-point one-loop diagram. The external momenta are
$q_1 = k_1' +p$, $q_2=-k_1'$, $q_3=-l'$, $q_4=l' -p$.
$p_\mu$ is a momentum regulator.}
\label{Twotree}
\end{center}
\end{figure}

At large $z$, the {\em integrand} behaves as
\be
\A_2^{\mathrm{1-loop}} =\frac {1} {2} \frac {\E1\cdot\E2}{l^2} + \frac {5} {2} \frac{\E1\cdot\E2}{k_1^2}\, \frac{ k_1\cdot l} {l^2} + \mathcal{O} \left( \frac{1}{z} \right) ~.
\ee
Upon integration over the loop momentum, the leading $\O(1)$ term becomes a linear combination of tadpole tensor integrals,
\be\label{eq9} I_{\mu_1\mu_2\dots} = \int \frac{d^{2\omega} l}{(4\pi)^{2\omega}} \, \frac{l_{\mu_1} l_{\mu_2} \cdots}{l^2} ~, \ee
which vanish. Therefore, we have no contribution from $z\to\infty$ and the entire contribution to the two-point diagram comes from the residue of the pole of the integrand at
\be\label{eq10} z = z_1 = \frac{(l-k_1)^2}{2\epsilon_1\cdot l} \ee
From Cauchy's theorem, we obtain for the integrand
\be\label{eq11} \mathcal{A}_2^{\mathrm{1-loop}} \Big|_{z=0} = -\frac{1}{z_1}\mathrm{Res}_{z\to z_1} \mathcal{A}_2^{\mathrm{1-loop}} + \dots \ee
where the dots represent contributions that vanish upon integration over the loop momentum.
Explicitly, for the {\em integral} we obtain
\be
A_{2}^{\mathrm{1-loop}}= +5 \E1^\mu \E2^\nu I_{\mu\nu}(k_1) + \frac {5} {2} \E1\cdot \E2 k_1^2 I(k_1)- \E1.\E2 k_1^\mu I_{\mu}(k_1) \label{TwoptIntegrand}
\ee
in terms of the two-point tensor integrals,
\be\label{eq13} I_{\mu_1\mu_2\dots} (k_1) = \int \frac{d^{2\omega} l}{(4\pi)^{2\omega}}\, \frac{l_{\mu_1} l_{\mu_2} \cdots}{l^2(l-k_1)^2 }~,
\ee
which is in agreement with the result of a direct calculation of the loop integral.
Evidently, the residue contributing to the loop amplitude is a four-point tree diagram contributing to the forward amplitude (see figure \ref{Twotree}),
\be\label{exp1}
 A_4^{\mathrm{tree}} (k_1', \epsilon_1 ; -k_1', \epsilon_2' ; l', \epsilon_3 ; -l' , \epsilon_4 ) \ee
where $\E2'$ is given by \eqref{eq2} with $z=z_1$ (defined in \eqref{eq10}), and we have defined
\be k_1' =k_1 + z_1\epsilon_1 \ \ , \ \  \ \ l' = l-k_1-z_1\epsilon_1 ~, \ee
to simplify the notation.
Two legs are on-shell, since $(l')^2 =0$.
For the two-point loop amplitude, we expect
\be\label{eq11n} A_2^{\mathrm{1-loop}} = \int  \frac{d^{2\omega} l}{(4\pi)^{2\omega}}\, \frac{1}{(l-k_1)^2} \sum_{\epsilon_3} A_4^{\mathrm{tree}} \Big|_{\epsilon_4 = \epsilon_3^*} \ee
However, the forward amplitude is singular. To regulate it, introduce a small momentum $p_\mu$ orthogonal to the polarization vector $\epsilon_2'$,
\be\label{eq11p} p\cdot \epsilon_2' = 0 \ee
and consider the amplitude
\be\label{eq11a} A_4^{\mathrm{tree}} \left( k_1', \epsilon_1 ; -k_1'-p, \epsilon_2'  ; l'+p , \epsilon_3 ; -l' , \epsilon_4 \right) \ee
in the limit $p_\mu\to 0$.

Since we are working with color-ordered amplitudes, to avoid ordering the legs carrying the loop momentum, we shall average over this amplitude and the one obtained by interchanging the two legs carrying the loop momentum.

The contribution of diagram \emph{(a)} of figure \ref{Twotree} is regular. In the limit $p_\mu \to 0$, we obtain
\be
\sum_{\E3} A_4^{\mathrm{tree}\, , \, (a)} \Big|_{\E4 = \E3^*} =\frac {\E1\cdot\E2' (\frac{5}{2} k_1^2 + k_1'\cdot l') +5 \E1\cdot l' \E2'\cdot l'+ \frac{5} {2} \E1.k_1'\E2'.l'} {l^2}
\ee
The contribution of the diagram \emph{(b)} of figure \ref{Twotree} is singular,
\be\label{eq17}
\sum_{\E3} A_4^{\mathrm{tree}\, , \, (b)}\Big|_{\E4 = \E3^*} =\frac {\E1\cdot\E2' (\frac{3}{2} p^2 + 3p\cdot k_1' - 3 p\cdot l' - 6 k_1'\cdot l') - 3 \E1\cdot l' \E2'\cdot p} {p^2}
\ee
Finally, the contribution of diagram \emph{(c)} is regular,
\be
\sum_{\E3}A_3^{\mathrm{tree}\, , \, (c)}\Big|_{\E4 = \E3^*} =- \frac{3} {2} \E1\cdot \E2'
\ee
The singular contribution is easily seen to vanish after removing the color ordering on the legs carrying the loop momentum. This is done by averaging with the expression obtained by replacing $l'\to -l'+p$ (or, equivalently, replacing $l \mapsto \frac{1}{2} p$ in \eqref{eq17}). Then the numerator on the right-hand side of \eqref{eq17} vanishes after using \eqref{eq11p}.

We obtain the finite forward tree amplitude
\be \sum_{\E3} A_4^{\mathrm{tree}}\Big|_{\E4 = \E3^*} =
 \frac {- 5\E1\cdot \E2 \E1\cdot l \frac{k_1\cdot l} {k_1^2} z_1 + \E1\cdot \E2 \E1\cdot l z_1 + \frac{3}{2} \E1\cdot \E2 k_1^2 + \E1\cdot \E2 k_1\cdot l + 5 \E1\cdot l \E2\cdot l}{l^2}
 \label{FinFwd}
 \ee
Substituting this expression in \eqref{eq11n}, we recover our earlier result \eqref{TwoptIntegrand} for the two-point one-loop amplitude.


The forward tree amplitude can also be obtained by applying the BCFW recursion relations. In fact, by an appropriate choice of complex momentum shifts, the singularity can be avoided and no need for a regulator arises.
Indeed, under the shift,
\be\label{eq20} k_1' \mapsto k_1' + w\E1 \ \ , \ \ \ \ -k_1' \mapsto -k_1' - w\E1 \ \ , \ \ \ \
\E2' \mapsto \E2'' \equiv \E2' - w \frac{\E1\cdot \E2}{k_1^2} k_1' \ee
the resulting amplitude vanishes as $w\to\infty$, and we obtain a pole at
\be\label{eq19} w = w_1 = - \frac{l^2}{2l\cdot\E1} \ee
The residue of the pole yields the entire four-point forward tree amplitude,
\bes\label{eqA4} A_4^{\mathrm{tree}} &=& \frac{1}{l^2}\sum_{\E3, \epsilon'} A_3^{\mathrm{tree}} (k_1'+w_1 \E1, \E1; -l'-w_1\E1, \epsilon'; l' , \E3) \nonumber\\
& & \times A_3^{\mathrm{tree}} (-k_1'-w_1\E1,\E2'' ; -l'+w_1\E1, \E3^*; l'+w_1\E1, {\epsilon'}^* ) \ees
which is a finite expression. After some straightforward algebra, we obtain
\be
A_4^{\mathrm{tree}}=\frac{\E1\cdot \E2 \left[ \frac{3} {2} w_1\E1\cdot l - \frac{7} {2} z_1 \E1\cdot l + \frac{5} {2} k_1^2 -  k_1\cdot l+  l^2 \right] - \frac{5} {2} \E1\cdot l \E2''\cdot k_1 + 5 \E1\cdot l \E2''\cdot l}{l^2}
\ee
Using the explicit expressions \eqref{eq10}, \eqref{eq19}, and \eqref{eq20} for $z_1$, $w_1$, and $\E2''$, respectively, we recover our earlier result \eqref{FinFwd}, which was obtained by a direct calculation using a regulator, up to terms which vanish upon integration over the loop momentum.

Thus, we have shown that an application of the BCFW recursion relations reduces the two-point loop amplitude to three-point tree amplitudes. Even though there are potential singularities from forward amplitudes, these were avoided by a judicious choice of complex momentum shifts.

\section{Three-point loop amplitude}
\label{sec:3}

Next we consider a three-point one-loop color-ordered amplitude
\be A_3^{\mathrm{1-loop}}(k_1, \E1 ; k_2, \E2 ; k_3, \E3) = \int \frac{d^{2\omega} l}{(4\pi)^{2\omega}} \mathcal{A}_3^{\mathrm{1-loop}}(k_1, \E1 ; k_2,\E2 ; k_3, \E3 )~, \ee
with $k_1+k_2+k_3 =0$. Two of the momenta, $k_1$ and $k_2$, will be on-shell. We shall keep the third momentum $k_3$ off shell to facilitate explicit calculations. This is necessary also for kinematical reasons, but $k_3^2 =0$ is allowed if momenta are complex, which is a case that will be useful for the calculation of higher-point amplitudes.

For the polarization vectors, we choose $\E1$ and $\E2$ such that $\E1\cdot k_i=0$ and $\E2\cdot k_i=0$, where $i=1,2,3$. This is always possible. Indeed, if $\E1\cdot k_2\ne 0$, then we may shift $\E1\mapsto \E1- \frac{\E1\cdot k_2}{k_1\cdot k_2} k_1$, and the new polarization vector satisfies $\E1\cdot k_i =0$. Similarly, we arrange $\E2\cdot k_i=0$. For the third polarization vector, since $k_3$ is off-shell, there are three independent polarizations. Notice that, since $\E3\cdot (k_1+k_2) =0$, they can be chosen as the set $\{ \E1, \E2, k_1-k_2 \}$.

\begin{figure}[ht]
\begin{center}
\includegraphics[width=12cm]{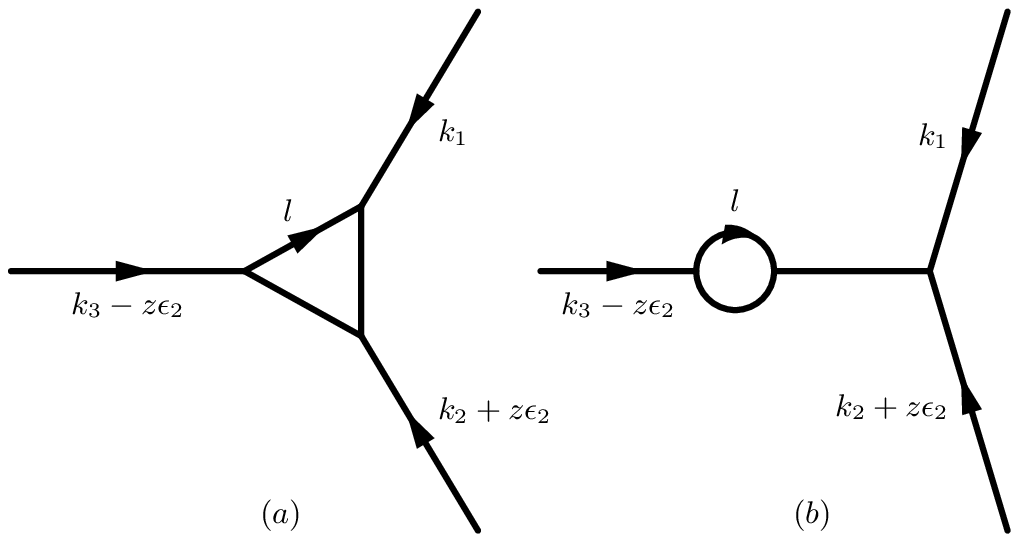}
\caption{Diagrams contributing to a three-point color-ordered one-loop amplitude.}
\label{fig:ThreePt}
\end{center}
\end{figure}

To apply the BCFW recursion relations, we shift
\be\label{eq20a} k_2 \mapsto k_2 + z \E2 \ , \ \ k_3 \mapsto k_3 - z \E2 \ , \ \ \E3 \mapsto \E3 + z \frac{\E2\cdot\E3}{k_3^2} k_3 ~. \ee
There are two diagrams that contribute to the amplitude (figure \ref{fig:ThreePt}) and we discuss them separately.

First we evaluate the triangle diagram {\em (a)} in figure \ref{fig:ThreePt} using the background gauge.
After the shift \eqref{eq20a}, the large $z$ behavior of the {\em integrand} is of the form
\bes
\mathcal{A}_3^{\mathrm{1-loop}\, , \, (a)} = \frac {1} {l^2(l+k_1)^2} & & \left[ -4 z {\E1}\cdot{\E2}
   {\E2}\cdot{\E3} +\frac{16 {\E1}\cdot l {\E2}\cdot {\E3} {k_2}\cdot l}{{k_3^2} }-\frac{4 {\E1}\cdot{\E2} {\E2}\cdot {\E3} {k_2}\cdot l}{
   {\E2}\cdot l }-   \frac{4 {\E1}\cdot {\E2} {\E2}\cdot {\E3} l\cdot {k_3}}{ {\E2}\cdot l } \right. \nonumber\\ & & \left. + \frac{2 {\E1}\cdot {\E2} {\E2}\cdot {\E3}}{{\E2}\cdot l } + 4 {\E1}\cdot {\E2}
   {\E3}\cdot {k_1} \right] + \mathcal{O} \left( \frac{1}{z} \right)
\ees
Evidently, it does not vanish, in general, as $z\to \infty$. Upon closer inspection, when $\E3= \E2$, or $\E3=k_1-k_2$, all terms except the last one at leading order ($\O(1)$) in the above expression vanish. The last term vanishes after integration over the loop momentum, because it is proportional to a two-point tensor integral \eqref{eq13} with $k_1^2=0$. Therefore, in the limit $z\to\infty$, there is no contribution.

If $\E3= \E1$, we need to interchange legs 1 and 2 before shifting the external momenta as in \eqref{eq20a}.

It turns out that the choices $\E3=\E1$ and $\E3=\E2$ yield vanishing amplitudes, so we shall concentrate on the polarization
\be\label{eq24} \E3 = k_1 - k_2 \ee
for which, as we just showed, there is no contribution to the diagram from the pole at $z\to\infty$.

It follows that the entire contribution to this diagram comes from the pole at
\be\label{eq25} z = z_1 \equiv - \frac{(l-k_3)^2}{2\epsilon_2\cdot l} \ee
Explicitly, for the integral we obtain
\be A_3^{\mathrm{1-loop}\, , \, (a)} \Big|_{z=0} = -8 {\E1}\cdot {\E2} {\E3}\cdot {k_1} k_3\cdot I(k_1,k_2)+4 {\E1}\cdot {\E2} {\E3}\cdot {k_1} I_\mu^\mu (k_1,k_2) - 4 {k_3^2} {\E1}\cdot{\E2} {\E3}\cdot I(k_1,k_2) +  16 {\E1^\mu} {\E2^\nu} {\E3^\lambda} I_{\mu\nu\lambda} (k_1,k_2) \label{3pta} \ee
in terms of three-point tensor integrals,
\be\label{eq24a} I_{\mu_1\mu_2\dots} = \int \frac{d^{2\omega}l}{(2\pi)^{2\omega}} \frac{l_{\mu_1}l_{\mu_2} \cdots}{l^2(l+k_1)^2(l+k_1+k_2)^2} \ee
After standard manipulations, we arrive at
\be
\label {3pteqI}
 A_3^{\mathrm{1-loop}\, , \, (a)} =\frac {1} {16\pi^2} {\E1}\cdot{\E2}{\E3}\cdot{k_1} \left( -\frac{20}{3 (2-\omega)} + \frac{40}{3} + \O(2-\omega) \right)~.
\ee
Next we compute diagram {\em (b)} in figure \ref{fig:ThreePt}
using the background gauge for the loop and the Gervais-Neveu gauge for the tree part of the diagram.

At large $z$, we obtain
\be
\A_3^{\mathrm{1-loop}\, , \, (b)} ={\E1}\cdot {\E2} {\E2}\cdot {\E3}\left[ -\frac{16  {k_2}\cdot l}{{k_3}^4 l^2}\, z -\frac{16 {k_1}\cdot l {k_2}\cdot l}{k_3^4 l^2 {\E2}\cdot l}-  \frac{16 {k_2}\cdot l {k_3}\cdot l}{k_3^4 l^2{\E2}\cdot l}+    \frac{4  {k_2}\cdot l}{{k_3^2} l^2 {\E2}\cdot l}+\frac{{k_2}\cdot l}{k_3^4 {\E2}\cdot l}\right] +
\frac{8 {\E1}\cdot {\E2} {\E3}\cdot l}{{k_3}^2 l^2} + \mathcal{O} \left( \frac{1}{z} \right) ~.
\ee
All $\O(z)$ and $\O(1)$ terms except the last one in the above expression vanish for the choice of polarization \eqref{eq24}. The last $\O(1)$ term also vanishes after
integration over the loop momentum (being proportional to a tadpole tensor integral \eqref{eq9}).

Proceeding as with the triangle diagram, we find that the residue of the pole
at $z=z_1$ \eqref{eq25} is the sole contribution. We obtain
\be
A_3^{\mathrm{1-loop}\, , \, (b)} = \frac{4\E1\cdot \E2 \E3^\mu}{k_3^2} \left[ - 2 k_3^2 {k_{1\mu}} I(k_3)  -  4 {k_1}^\nu I_{\mu\nu} (k_3) -  4 {k_3}^\nu I_{\mu\nu} (k_3)  + 2 I_{\mu\nu}^{\ \ \ \ \nu} (k_3) +  k_3^2 I_\mu (k_3) \right]
\label{3ptb}\ee
written in terms of two-point tensor integrals \eqref{eq13}.

After integrating over the loop momentum, we arrive at
\begin {equation}
\label {3pteqII}
A_3^{\mathrm{1-loop}\, , \, (b)}=\frac {1} {16\pi^2}{\E1}\cdot{\E2}{\E3}\cdot{k_1} \left(\frac{20}{3 (2-\omega) } -12 + \mathcal{O} (2-\omega)  \right).
\end {equation}
Adding the contributions of the two diagrams, \eqref{3pteqI} and \eqref{3pteqII}, we obtain a finite three-point one-loop amplitude,
\begin {equation}
A_3^{\mathrm{1-loop}}=A_3^{\mathrm{1-loop}\, , \, (a)}+ A_3^{\mathrm{1-loop}\, , \, (b)}= \frac {1} {12\pi^2} {\E1}\cdot{\E2}{\E3}\cdot{k_1} ~,
\label{tot3pt}
\end {equation}
as expected \cite{Bern:1995ix}.

Recall that this is valid for a choice of polarization vectors $\E1$ and $\E2$ obeying $\E1\cdot k_i = \E2\cdot k_i =0$ ($i=1,2,3$). It is easily generalized to arbitrary polarization vectors,
\begin {equation}\label{eq31}
A_3^{\mathrm{1-loop}}=A_3^{\mathrm{1-loop}\, , \, (a)}+ A_3^{\mathrm{1-loop}\, , \, (b)}= \frac {1} {12\pi^2} A_3^{\mathrm{tree}} \ , \ \ A_3^{\mathrm{tree}} = {\E1}\cdot{\E2}{\E3}\cdot{k_1} + \E2\cdot \E3 \E1\cdot k_2 + \E3\cdot \E1 \E2\cdot k_3 ~.
\end {equation}
This form is also valid in the limit in which all three legs are on shell ($k_i^2 = 0$, $i=1,2,3$), which is kinematically allowed if the momenta are complex, and will be useful in the calculation of higher-order diagrams.
On shell $k_3$ has two polarizations which can be chosen as the set of null vectors $\{ k_1-k_2 \ , \ \E2\cdot k_1 \E1 - \E1\cdot k_2 \E2 - \E1\cdot\E2 \frac{k_1-k_2}{2} \}$. Once again, only polarizations that have non-vanishing components along $\E3 = k_1-k_2$ give non-vanishing amplitudes.

\begin{figure}[ht!]
\begin{center}
\includegraphics[width=15cm]{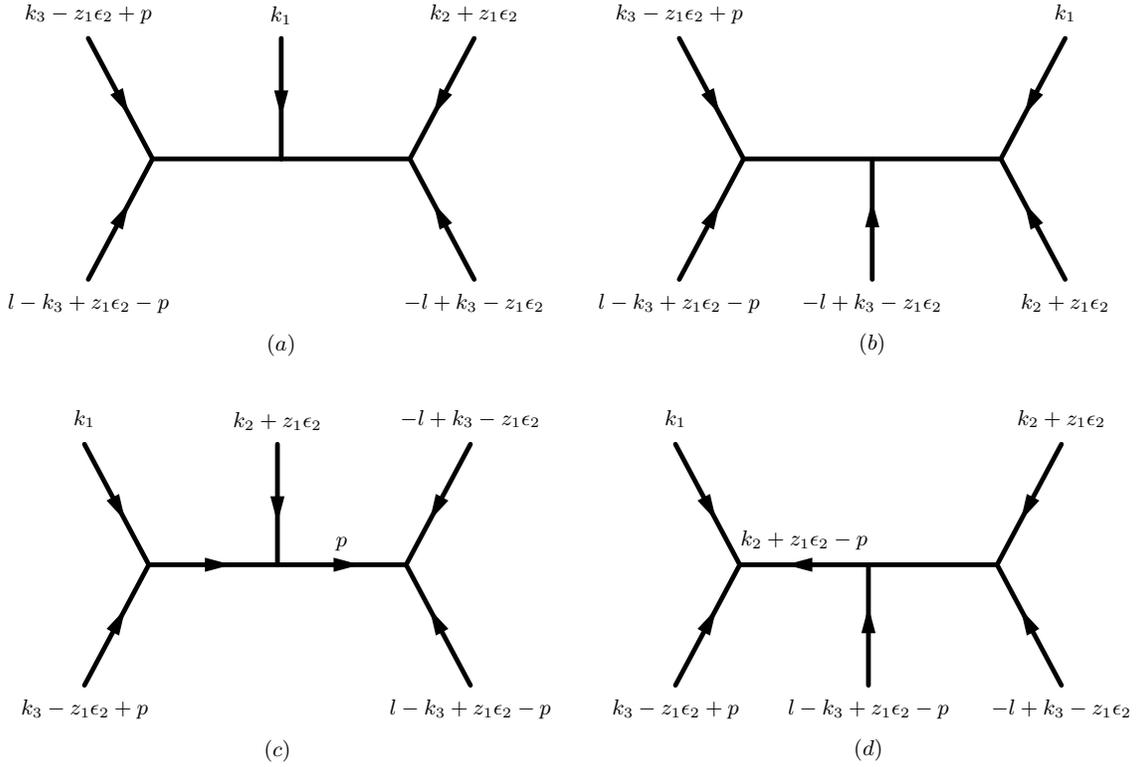}
\caption{Some of the five-point tree diagrams contributing to a three-point color-ordered one-loop amplitude. $p_\mu$ is a momentum regulator.}
\label{Threetree}
\end{center}
\end{figure}

Evidently, the residue contributing to the loop amplitude consists of two five-point tree diagrams contributing to the forward amplitude (diagrams {\em (a)} and {\em (b)} in figure \ref{Threetree}),
\be\label{eqA5} A_5^{\mathrm{tree}} ( k_2+z_1\epsilon_2, \epsilon_2 ; k_1, \epsilon_1; k_3-z_1\epsilon_2, \epsilon_3 ; l-k_3+z_1\epsilon_2 , \epsilon_4 ; -l+k_3-z_1\epsilon_2 , \epsilon_5 ) \ee
with $z_1$ given by \eqref{eq25}. All legs are on-shell, but we shall keep the momentum $k_3$ off shell for convenience, taking the limit $k_3^2\to 0$ at the end of the day. The contributions of the first two diagrams in figure \ref{Threetree}, $A_5^{\mathrm{tree}\, , \, (a)}$ and $A_5^{\mathrm{tree}\, , \, (b)}$, respectively, match our earlier result after we identify $\epsilon_5= \epsilon_4^*$ and sum over the polarization vectors $\epsilon_4$. We conclude
\be A_3^{\mathrm{1-loop}} = \int  \frac{d^{2\omega} l}{(4\pi)^{2\omega}}\, \frac{1}{(l-k_3)^2} \sum_{\epsilon_4} \Big( A_5^{\mathrm{tree}\, , \, (a)} + A_5^{\mathrm{tree}\, , \, (b)} \Big) \Big|_{\epsilon_5 = \epsilon_4^*} \ee

However, the forward tree amplitude is singular. To regulate it, introduce a small momentum $p_\mu$ and consider the amplitude with shifted legs $k_3-z_1\E2\mapsto k_3-z_1\E2 + p$, $l-k_3+z_1\E2\mapsto l-k_3+z_1\E2 -p$ (figure \ref{Threetree}),
in the limit $p_\mu\to 0$.
As with the two-point loop amplitude, it can be checked that the singular terms do not contribute after integration over the loop momentum. We conclude
\be\label{eq37} A_3^{\mathrm{1-loop}} = \int  \frac{d^{2\omega} l}{(4\pi)^{2\omega}}\, \frac{1}{(l-k_3)^2} \sum_{\epsilon_4} A_5^{\mathrm{tree}}\Big|_{\epsilon_5 = \epsilon_4^*} \ee
The calculation of the forward amplitude $A_5^{\mathrm{tree}}$ can be done by applying the BCFW recursion relations. By appropriate shifts of momenta, it can thus be reduced to three-point tree amplitudes avoiding the singularities. Indeed, let us shift
\be  k_1\mapsto  k_1 + w \E2 \ , \ \ \ k_2 +z_1 \E2 \mapsto k_2+z_1 \E2 - w  \E2 \ee
The contribution from $w\to\infty$ is easily seen to vanish. There is a pole at
\be w=w_1 = -\frac{ (l+k_1)^2} {2l\cdot\E2}~ .\ee
Its residue gives the entire five-point tree amplitude \eqref{eqA5},
\bes \frac{\mathrm{Res}_{w\to w_1}}{w_1} &=& \sum_{\E4} A_5^{\mathrm{tree}}\Big|_{\epsilon_5 = \epsilon_4^*} \nonumber\\
&=&
\frac{1}{(l+k_1)^2} \sum_{\E4,\epsilon'} A_3^{\mathrm{tree}} (-l+k_3-z_1\E2, \E4; k_2 + z_1 \E2-w_1 \E2,\E2; l+k_1+w_1 \E2, \epsilon)
\nonumber\\ & & \ \ \ \ \ \ \ \ \ \ \ \ \times A_4^{\mathrm{tree}} (l-k_3+ z_1 \E2,{\E4}^* ; k_3-z_1 \E2, \E3 ;k_1+w_1 \E2, \E1  ; -l -k_1-w_1 \E2 ,\epsilon^* ).\ees
The four point amplitude in the above expression is a forward amplitude. It can be reduced to a finite expression involving three-point tree amplitudes, as before (see discussion in the case of the two-point loop amplitude leading to eq.\ \eqref{eqA4}).
After some straightforward algebra, we arrive at the finite expression
\be
\sum_{\E4} A_5^{\mathrm{tree}}\Big|_{\epsilon_5 = \epsilon_4^*}  = \frac{4\N}{l^2(l+k_1)^2(l-k_3)^2} \ee
where
\be\label{eq41} \N = \frac{4 {\E1}\cdot {\E2} {\E3}\cdot l ( l+ {k_1})^2
   {k_2}\cdot l}{k_3^2}-2 {\E1}\cdot {\E2}
   {\E3}\cdot {k_1} ( l-{k_3})^2+
   k_3^2 {\E1}\cdot {\E2} {\E3}\cdot (l+{k_1})-2
   {\E1}\cdot {\E2} {\E3}\cdot {k_1} ( l+ {k_1})^2l-4 {\E1}\cdot l
   {\E2}\cdot l {\E3}\cdot l
\ee
which indeed yields the sum of \eqref{3pta} and \eqref{3ptb} (via \eqref{eq37}), and therefore the correct (finite) value of the three-point loop amplitude \eqref{tot3pt}.

\section{Four-point loop amplitude}
\label{sec:4}

In this section, we consider the four-point color-ordered one-loop amplitude,
\be\label{eq39} A_4^{\mathrm{1-loop}}(k_1,\E1 ; k_2, \E2 ; k_3, \E3; k_4, \E4) = \int \frac{d^{2\omega} l}{(4\pi)^{2\omega}} \mathcal{A}_4^{\mathrm{1-loop}}(k_1,\E1 ; k_2, \E2 ; k_3, \E3; k_4, \E4) \ee
where $k_1+k_2+k_3+k_4 =0$ and all momenta are on shell ($k_1^2=k_2^2=k_3^2=k_4^2=0$).

It suffices to consider amplitudes in which
\be\label{eqE1E2} \E1 = \E2 \ee
This is because they form a basis: all amplitudes can be expressed as linear combinations of amplitudes with two identical polarization vectors.
To see this, first recall that for general momenta $k_1$ and $k_2$, the corresponding polarization vectors can be chosen to be common to both. Indeed, if $\E1\cdot k_2\ne 0$, then by shifting $\E1\mapsto \E1 - \frac{\E1\cdot k_2}{k_1\cdot k_2} k_2$, we satisfy $\E1\cdot k_2=0$ (in addition to $\E1\cdot k_1=0$). There are two linearly independent choices for $\E1$ obeying $\E1\cdot k_2 =\E1\cdot k_1 =0$. Similarly, we have two linearly independent choices of $\E3$ such that $\E3\cdot k_2 =\E3 \cdot k_3 =0$. Then a basis for the polarization vector $\E2$ can be $\{ \E1, \E3\}$. Thus, we need only consider amplitudes with $\E2=\E1$ or $\E2=\E3$. Without loss of generality, we adopt \eqref{eqE1E2}.

To apply the BCFW recursion relations, we shall shift the two adjacent legs,
\be\label{eq43} k_1 \mapsto k_1 + z \E1 \ , \ \ k_2 \mapsto k_2 - z \E1~. \ee
An explicit calculation shows that for polarization vectors obeying \eqref{eqE1E2}, the \emph{integrand} of the four-point one-loop amplitude \eqref{eq39} vanishes in the limit $z\to\infty$,
\be \A_4^{\mathrm{1-loop}}(\E1 , k_1 +z \E1; \E2 , k_2 -z\E1; \epsilon_3, k_3; \E4, k_4) = \mathcal{O} \left( \frac{1}{z} \right) \ee
Therefore, only the poles contribute to the amplitude. To calculate their residues,
it is advantageous to consider the basis for the remaining polarization vectors, $\E3$ and $\E4$,
\be \E3 = \left\{ \E1 - \frac{k_3\cdot \E1}{k_3\cdot k_1}  k_1 \ , \ \E2 - \frac{k_3\cdot \E2}{k_3\cdot k_2} k_2 \right\} \ , \ \ \E4 = \left\{ \E1 - \frac{k_4\cdot \E1}{k_4\cdot k_1}  k_1 \ , \ \E2- \frac{k_4\cdot \E2}{k_4\cdot k_2} k_2 \right\} \ee

\subsection{Choice (A) of polarization vectors}

First, consider the case
\be\label{eq42} \mathbf{(A) \ \ : } \ \ \ \ \ \ \ \E3 = \E1 -\frac{k_3\cdot \E1}{k_3\cdot k_1} k_1 \ , \ \ \E4 =  \E1 -\frac{k_4\cdot \E1}{k_4\cdot k_1} k_1 \ee
Notice that with this choice of polarization vectors, the corresponding four-point tree diagram vanishes.

\begin{figure}[t]
\begin{center}
\includegraphics[width=15cm]{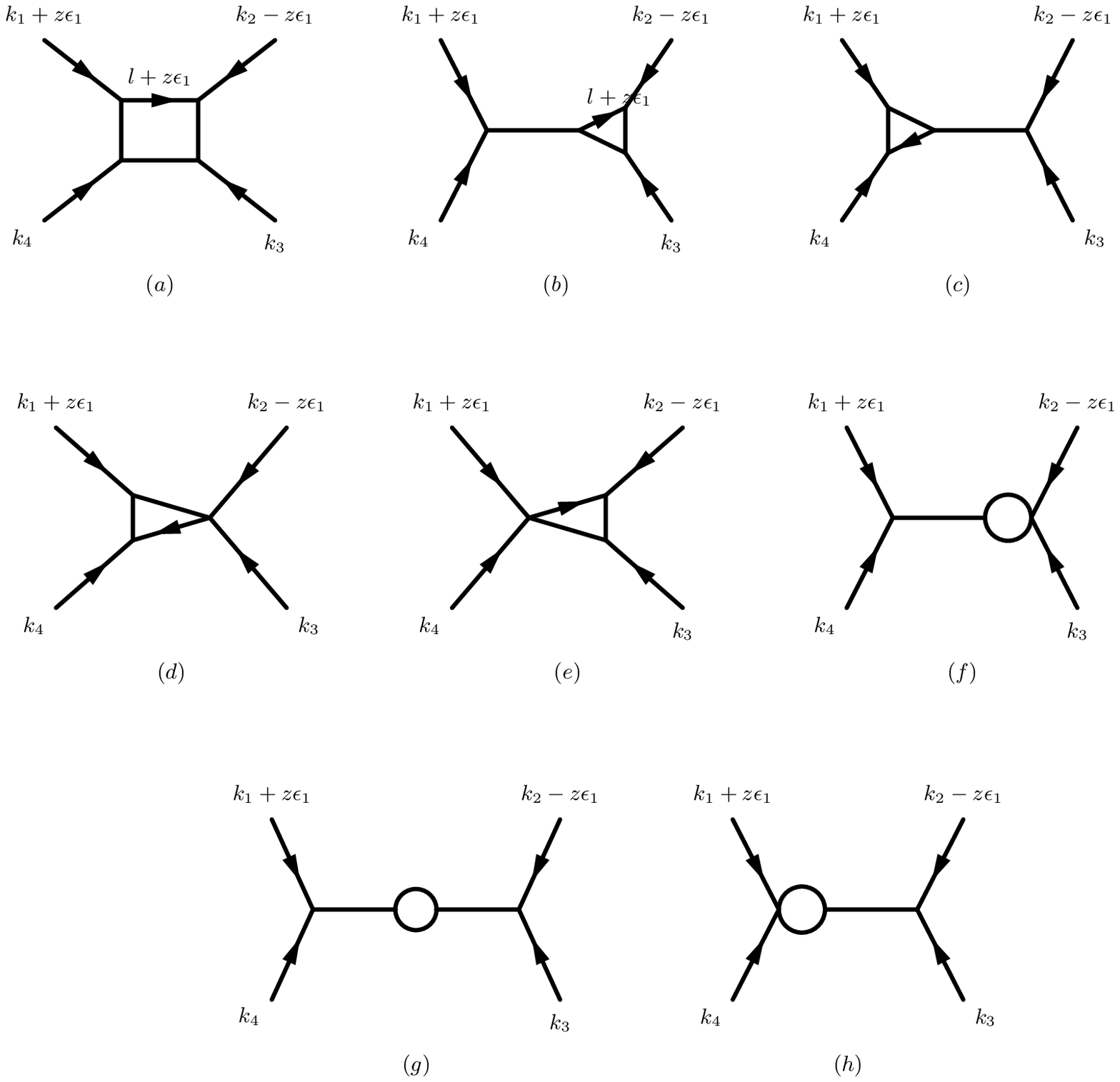}
\caption{Diagrams contributing to a four-point color-ordered one-loop amplitude.}
\label{fig:Box}
\end{center}
\end{figure}

The entire contribution to the box diagram in figure \ref{fig:Box} comes from the pole at
\be\label{eq48} z = z_1 = - \frac{l^2}{2\E1\cdot l} \ee
Explicitly,
\be A_4^{\mathrm{1-loop}\, ,\, (a)} \Big|_{z\to z_1} = 16\E1^\mu \E1^\nu \left[ \alpha^{\rho\sigma} I_{\mu\nu\rho\sigma} (k_2,k_3,k_4) + \beta^{\rho} I_{\mu\nu\rho} (k_2,k_3,k_4) \right] \ee
written in terms of the four-point tensor integrals,
\be\label{eq50} I_{\mu_1\mu_2\dots} (k_2, k_3, k_4) = \int \frac{d^{2\omega} l}{(2\pi)^{2\omega}} \frac{l_{\mu_1}l_{\mu_2} \cdots}{l^2 (l+k_2)^2 (l+k_2+ k_3)^2 (l+k_2+k_3+k_4)^2} \ee
where
\bes\label{eq51} \alpha^{\rho\sigma} &=& - \E1^\rho\E1^\sigma +  \frac{{\E1}\cdot{k_3} {(k_2-k_1)}\cdot{k_3} \E1^\rho k_1^\sigma  +  ({\E1}\cdot{k_3})^2 k_1^\rho k_1^\sigma }{{k_1}\cdot{k_3} {k_2}\cdot{k_3}}\nonumber\\
\beta^\rho &=&  {\E1}\cdot {k_3} \frac{{k_1}\cdot {k_2}}{{k_1}\cdot{k_3} } \E4^\rho \ees
After we integrate over the loop momentum, we obtain a finite expression,
\be
\label{4eqI}
A_4^{\mathrm{1-loop}\, , \, (a)}\Big|_{z\to z_1} = -\frac {1} {24 \pi^2} \frac {(\E1\cdot k_3)^4 k_1\cdot k_2}{ k_1\cdot k_3 (k_2\cdot k_3)^2}~.
\ee
There is one more diagram that contributes to this amplitude (diagram {\em (b)} in figure \ref{fig:Box}). The other diagrams vanish for the choice of polarization vectors under consideration (eqs.~\eqref{eqE1E2} and \eqref{eq42}).

Diagram {\em (b)} in fig.\ \ref{fig:Box} has two poles, one given by \eqref{eq48}, and a new pole at
\be\label{eq53} z = z_2 = - \frac{k_2\cdot k_3}{\E1\cdot k_3} \ee
The residue of the pole \eqref{eq48} gives a contribution to the amplitude,
\be A_4^{\mathrm{1-loop}\, ,\, (b)}\Big|_{z\to z_1} = 16\E1^\mu \E1^\nu \left[ \alpha^{\rho\sigma} I_{\mu\nu\rho\sigma} (k_2,k_3,k_4') + \beta^{\rho} I_{\mu\nu\rho} (k_2,k_3,k_4') \right] \ee
where we introduced the on-shell momentum (it is easy to see that $k_4^{\prime 2} =0$),
\be k_4' = \frac{k_2\cdot k_3}{\E1\cdot k_3} \E1 - k_2 -k_3 \ee
and the coefficients $\alpha^{\rho\sigma}$ and $\beta^{\rho}$ are as before (eq.\ \eqref{eq51}). It is easily seen to vanish (by a direct calculation, or, e.g., by replacing $k_1 \mapsto z_2 \E1$ in \eqref{4eqI}),
\be A_4^{\mathrm{1-loop}\, ,\, (b)}\Big|_{z\to z_1} = 0 ~. \ee
Therefore \eqref{4eqI} is the entire contribution of the pole \eqref{eq48}.

Working as above with the second pole \eqref{eq53}, after some straightforward algebra we find that
the residue of the pole \eqref{eq53} gives a finite contribution to the amplitude,
\be\label{4eqII}
A_4^{\mathrm{1-loop}\, ,\, (b)} \Big|_{z\to z_2}
=\frac {(\E1\cdot k_3)^4 (k_1\cdot k_2)^2}{24 \pi^2 k_1\cdot k_3 (k_2\cdot k_3)^3}~.
\ee
Notice that each pole contribution can be written as a single term and the two poles lead to different kinematical expressions.

Combining the contribution of the two poles, \eqref{4eqI} and \eqref{4eqII}, we obtain the four-point amplitude
\be
\label{FinSum}
A_4^{\mathrm{1-loop}} = \frac {(\E1\cdot k_3)^4 k_1\cdot k_2 (k_1-k_3)\cdot k_2}{24 \pi^2 k_1\cdot k_3 (k_2\cdot k_3)^3}~,
\ee
which is the same expression (with appropriate identifications) as in \cite{Bern:1991aq}.

\begin{figure}[ht]
\begin{center}
\includegraphics[width=9cm]{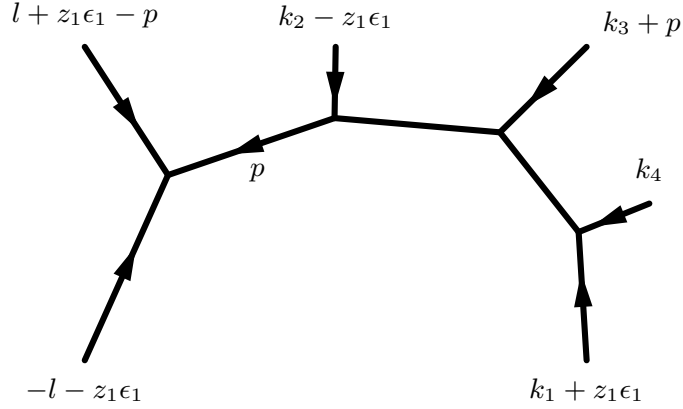}
\caption{A six-point tree diagram that contributes to the four-point color-ordered one-loop amplitude. $p_\mu$ is a momentum regulator.}
\label {FourTree1}
\end{center}
\end{figure}

The residue at $z=z_1$ \eqref{eq48} can be expressed in terms of a six-point forward tree amplitude. As in the case of a three-point one-loop amplitude, we can introduce a momentum regulator $p_\mu$ by shifting the legs
$l+z_1\E1\mapsto l+z_1\E1-p$, $k_3\mapsto k_3+p$ (see figure \ref{FourTree1}).
An explicit calculation shows that singularities of the forward amplitude do not contribute (in the limit $p_\mu \to 0$) after integration over the loop momentum.

Thus, the contribution to the pole at $z=z_1$ can be written as
\be\label{eqA6} A_4^{\mathrm{1-loop}}\Big|_{z\to z_1} = \int  \frac{d^{2\omega} l}{(4\pi)^{2\omega}}\, \frac{1}{l^2} \sum_{\E5} A_6^{\mathrm{tree}}\Big|_{\E6 = \E5^*} \ee
As with the five-point tree amplitude involved in the calculation of a three-point loop amplitude,
the six-point amplitude can be reduced to lower-point amplitudes by a judicious application of the BCFW recursion relations.
It is convenient to
shift
\be\label{eqshA} k_1+z_1\E1 \mapsto k_1+z_1\E1 + w \E4 \ \ , \ \ \ \ k_4 \mapsto k_4 - w \E4~. \ee
There is no shift in the polarization vectors, because $\E1\cdot \E4 =0$ because of \eqref{eq42}.
One can easily check that the amplitude vanishes in the limit $w\to\infty$. There is a pole at
\be
\label{polew1}
w=w_1 = \frac{(l-k_1)^2}{2 \E4 \cdot l}
\ee
The corresponding residue is given by
\bes \frac{\mathrm{Res}_{w\to w_1}}{w_1} &=& \sum_{\E5} A_6^{\mathrm{tree}} \Big|_{\E6 = \E5^*} \nonumber\\
&=& \frac{1}{(l-k_1)^2}
\sum_{\E5, \epsilon'} A_5^{\mathrm{tree}} ( l+z_1\E1, \E5^* ; k_2-z_1 \E1, \E1; k_3, \E3 ; k_4- w_1 \E4, \E4;
 -l+k_1+ w_1 \E4, \epsilon) \nonumber\\
& & \ \ \ \ \ \ \ \ \ \ \ \ \ \ \times A_3^{\mathrm{tree}} (l-k_1- w_1 \E4,\epsilon^\prime; k_1+z_1 \E1+ w_1 \E4,\E4 ;  -l-z_1\E1, \E5)
\ees
The five-point tree amplitude is a forward amplitude containing potential singularities. However, it can be calculated in the same way as the five-point forward amplitude encountered in the calculation of the three-point loop amplitude (see eqs.\ \eqref{eq37} through \eqref{eq41}). Thus, by a repeated application of the BCFW recursion relations, it is reduced to on-shell three-point amplitudes. After some algebra, and using \eqref{eqA6}, we obtain agreement with our earlier result \eqref{4eqI}, which was obtained by a direct diagrammatic calculation.

Turning to the other pole that contributes to the amplitude, at $z=z_2$, we obtain the residue
\be\label{eqpole2} \A_4^{\mathrm{1-loop}}\Big|_{z\to z_2} = \frac{1}{k_2\cdot k_3} \sum_{\epsilon'} A_3^{\mathrm{1-loop}} (k_2-z_2\E1,\E1; k_3, \E3; -k_2-k_3+z_2\E1, \epsilon')
A_3^{\mathrm{tree}} (k_1+z_2\E1,\E1;-k_1-k_4-z_2\E1,\epsilon' ; k_4, \E4) ~. \ee
It is already written in terms of on-shell amplitudes with no singularities. Using our earlier results on three-point amplitudes, and integrating over the loop momentum, after some algebra, one can show that the contribution of the second pole \eqref{eqpole2} agrees with our earlier result \eqref{4eqII} obtained by a direct diagrammatic calculation.

Thus, we have shown that the four-point one-loop amplitude with the choice of polarization vectors \eqref{eq42} can be expressed in terms of three-point on-shell tree-amplitudes and a three-point one-loop on-shell amplitude \eqref{eq31}. The latter also reduces to three-point tree-amplitudes, as was shown in the previous section.

\subsection{Choice (B) of polarization vectors}

Next, we consider the case of polarization vectors
\be\label{eq42b} \mathbf{(B) \ \ : } \ \ \ \ \ \ \ \E3 = \E1 -\frac{k_3\cdot \E1}{k_3\cdot k_1} k_1 \ , \ \ \E4 =  \E2 -\frac{k_4\cdot \E2}{k_4\cdot k_2} k_2 \ee
Unlike with the previous choice \eqref{eq42}, the corresponding four-point tree diagram is non-vanishing,
\be
A_4^{\mathrm{tree}\, ,\, \mathbf{(B)}} =- \frac {(\E1\cdot k_3)^4 k_1\cdot k_2} {(k_1\cdot k_3)^2 k_2\cdot k_3}~.
\ee
One obtains a simple expression because only the $t$-channel contributes to the color-ordered amplitude.

For the loop amplitude, we obtain eight non vanishing graphs which contribute for our choice of basis \eqref{eq42b} shown in figure \ref{fig:Box}.
A direct calculation shows that the pole at $z=z_2$ \eqref{eq53} gives a vanishing contribution. This is confirmed by an application of the BCFW recursion relations (eq.\ \eqref{eqpole2}).
Therefore, the amplitude is determined solely by the pole at $z=z_1$ \eqref{eq48}. A calculation of the residue of the pole, using diagrams as before, leads to an expression which is in agreement with the one obtained by a direct diagrammatic calculation.
After integrating over the loop momentum, we obtain a divergent expression,
\bes
\label{Int4pt}
A_4^{\mathrm{1-loop}}&=& \frac {1}{8 \pi^2} A_4^{\mathrm{tree} \, , \, \mathbf{(B)}} \frac{\Gamma^2 (\omega -1) \Gamma(3-\omega)}{ \Gamma(2\omega -3)} \left( \frac{4\pi\mu^2}{s} \right)^{2-\omega} \nonumber\\
& & \times \left[ -\frac{2}{(2-\omega)^2} - \frac{1}{2-\omega} \left( \frac{11}{3} - 2\ln \frac{t}{s} \right) + \frac{11}{6} \ln\frac{\mu^2t}{s^2} + \frac{\pi^2}{2} - \frac{32}{9}+ \mathcal{O}(2-\omega)\right] ~,
\ees
where $s=2k_1\cdot k_2$, $t = 2k_2\cdot k_3$. This expression agrees with the ones derived in \cite{Bern:1991aq} (see also \cite{Ellis:1986aq}) with appropriate kinematical identifications, after setting the arbitrary momentum scale $Q^2 =s$.

The contribution to the pole at $z=z_1$ can be written in terms of a six-point forward tree amplitude as in \eqref{eqA6}.
The latter can be reduced to lower-point amplitudes by a judicious application of the BCFW recursion relations (avoiding the potential singularities).
To this end, instead of the shift \eqref{eqshA}, it is convenient to shift
\be\label{eqshB} k_3 \mapsto k_3+ w q \ \ , \ \ \ \ k_4 \mapsto k_4 - w q \ \ , \ \ \ \ q = \E1 - \frac{\E1\cdot k_3 }{k_1\cdot k_3} (k_1+k_3)~.\ee
There is no shift in the polarization vectors, because $\E3\cdot q = \E4\cdot q =0$, where we used \eqref{eqE1E2} and \eqref{eq42b}.
In fact $\E{i} -q$ is along the direction of the corresponding momentum $k_i$ ($i=3,4$). Since the amplitude is on shell, we could replace both polarization vectors $\E3$ and $\E4$ by $q$, to simplify the calculation.

One can easily check that the amplitude vanishes in the limit $w\to\infty$. There is a pole at
\be
\label{polew1B}
w=w_1 = - \frac{(l+k_2+k_3)^2}{2 q \cdot (l+k_2)}
\ee
The corresponding residue is given by
\bes \frac{\mathrm{Res}_{w\to w_1}}{w_1} &=& \sum_{\E5} A_6^{\mathrm{tree}} \Big|_{\E6 = \E5^*} \nonumber\\
&=&  \frac{1}{2 (l+k_2) \cdot q} \sum_{\E5, \epsilon'} A_4^{\mathrm{tree}} ( l+z_1\E1, \E5^* ; k_2-z_1 \E1, \E1; k_3+wq, q ; -l-k_2-k_3-w q, \epsilon^{\prime *}) \nonumber\\
& & \ \ \ \ \ \ \ \ \ \ \ \ \ \ \ \ \ \ \ \  \times A_4^{\mathrm{tree}} (l+k_2+k_3+w q,\epsilon^\prime; k_4-w q,q ; k_1+z_1\E1, \E1; -l-z_1\E1, \E5)
\ees
The two four-point tree amplitudes are on-shell amplitudes and can be reduced to three-point amplitudes by an application of the BCFW recursion relations. Thus, by a repeated application of the BCFW recursion relations, the six-point amplitude is reduced to on-shell three-point amplitudes. After some algebra, and using \eqref{eqA6}, we obtain agreement with our earlier result \eqref{Int4pt}, which was obtained by a direct diagrammatic calculation.

The remaining two choices of polarization vectors can be tackled similarly and will not be discussed explicitly here.

Summarizing, we have shown that four-point one-loop amplitudes can be expressed in terms of three-point on-shell tree-amplitudes.

\section{Higher-point loop amplitudes}
\label{sec:5}

The calculation of the four-point color-ordered one-loop amplidute can be straightforwardly generalized to high-point amplitudes,
\be A_n^{\mathrm{1-loop}}\left( \{ k_i , \E{i} \} \right) = \int \frac{d^{2\omega} l}{(4\pi)^{2\omega}} \A_n^{\mathrm{1-loop}}\left( \{ k_i , \E{i} \} \right) \ee
As explained in section \ref{sec:4}, it suffices to consider amplitudes with two identical polarization vectors. Without loss of generality, we shall choose \eqref{eqE1E2} for the adjacent legs with momenta $k_1, k_2$.

\begin{figure}[ht]
\begin{center}

\includegraphics[width=15cm]{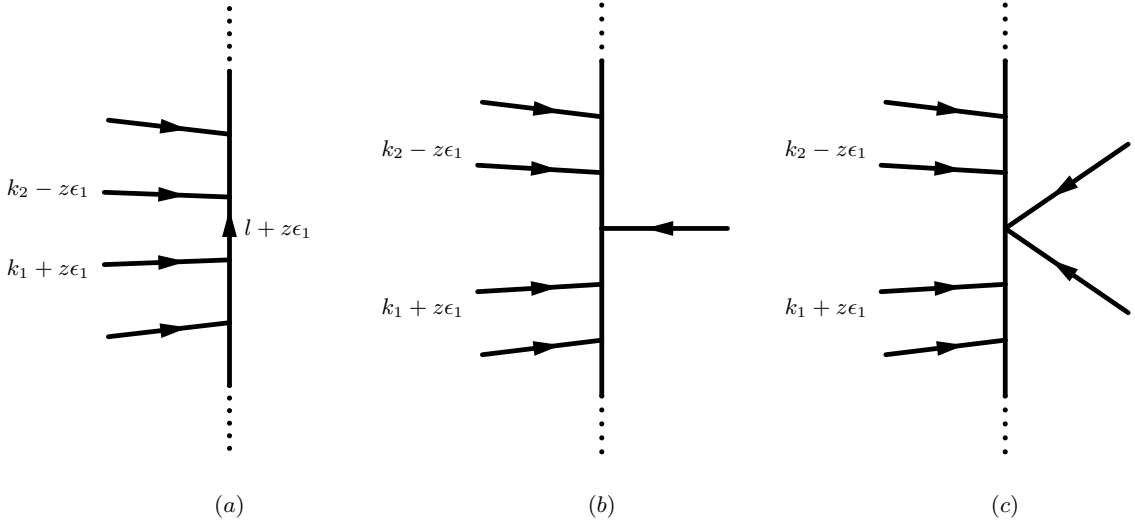}
\caption{Diagrams contributing to higher-point amplitudes.}
\label {npoint}
\end{center}
\end{figure}

To apply the BCFW recursion relations, we shift the momenta $k_1, k_2$ as in \eqref{eq43}.
Using the Ward identity,
\be A_n^{\mathrm{1-loop}}(k_1+ z\E1 , k_1 +z\E1 ; \dots ) = 0 \ee
we deduce
\be\label{eq45h} A_n^{\mathrm{1-loop}}(k_1 +z \E1 , \E1; k_2 -z\E1 , \E1; k_3 , \E3; \dots ; k_n, \E{n}) = - \frac{1}{z} A_n^{\mathrm{1-loop}}(k_1 +z\E1 , k_1; k_2-z\E1 , \E1; k_3, \E3 ; \dots ; k_n , \E{n}) \ee
It is easy to see that the amplitude on the right-hand side of \eqref{eq45h} has a finite limit as $z\to\infty$. Indeed, e.g., diagram {\em (a)} in figure \ref{npoint}, is a rational function of $z$. There are two  $\mathcal{O}(z)$ vertices that contribute to the numerator, and one  $\mathcal{O}(z)$ propagator that contributes to the denominator. The  $\mathcal{O}(z)$ contribution is the leading term,

\be
\A_n^{\mathrm{1-loop}\ , \ (a)}\Big|_{\E1 = k_1} = \frac{\cdots k_1^{\mu_1}\left[ -\eta_{\rho \nu} \epsilon_{1\mu_1}- 2 \eta_{\nu \mu_1} \epsilon_{1\rho}+  2 \eta_{\mu_1 \rho} \epsilon_{1\nu}\right]  \eta^{\rho\sigma} \E1^{\mu_2}\left[ -\eta_{\sigma \lambda} \epsilon_{1\mu_2}+ 2 \eta_{\lambda \mu_2} \epsilon_{1\sigma}-  2 \eta_{\mu_2 \sigma} \epsilon_{1\lambda} \right]\cdots }{ 2 \E1\cdot l} z + \O(1)
\ee
Evidently, the numerator of the leading $\O(z)$ term vanishes, showing that the contribution of this diagram is $\O(1)$. Similarly, one can show that the $\O(z)$ terms in all other diagrams, such as {\em (b)} and {\em (c)} in figure \ref{npoint} vanish, therefore all diagrams contributing to the amplitude on the right-hand side of \eqref{eq45h} (with $\E1 = k_1$) are finite in the limit $z\to\infty$, and the amplitude we are interested in (left-hand side of \eqref{eq45h}) is
\be \A_n^{\mathrm{1-loop}}(k_1 +z \E1, \E1; k_2 -z\E1, \E1; k_3, \E3; \dots ; k_n, \E{n}) = \mathcal{O} \left( \frac{1}{z} \right) \ee
Thus, only poles contribute to the integrand. The pole in the one-particle irreducible part of the amplitude has a residue which is a forward tree amplitude with $n+2$ legs. The extra two legs have momenta $\pm (l + z_1\E1)$ and corresponding polarization vectors $\E{{n+1}}$ and $\E{{n+2}}$, with $\E{{n+2}} = \E{{n+1}}^*$ and we need to sum over $\E{{n+1}}$. Additional poles exist on propagators which lead to a factorized amplitude when cut. Putting these together, we obtain for the loop amplitude
\bes\label{eq76a} A_n^{\mathrm{1-loop}} = & & \int \frac{d^{2\omega} l}{(2\pi)^{2\omega}} \frac{1}{l^2} \sum_{\E{{n+1}}} A_{n+2}^{\mathrm{tree}} \Big|_{\E{{n+2}} =\E{{n+1}}^*} \nonumber\\
& & + \sum_I  \frac{1}{(\sum_{i\in I}k_i)^2} \sum_{\epsilon} A_m^{\mathrm{1-loop}} \left( \{ k_i , \E{i} \}_{i\in I}; -\sum_{i\in I} k_i , \epsilon' \right) A_{n-m}^{\mathrm{tree}} \left( -\sum_{j\in J} k_j , \epsilon^{\prime*} ; \{ k_j , \E{j} \}_{j\in J} \right) \ees
where the second term consists of the contributions of the residues of the poles $z=z_I$, where
\be z_I = \frac{K^2}{2\E1\cdot K} \ , \ \ K = \sum_{i\in I} k_i ~, \ee
and we sum over all poles, i.e., all possible partitions of the set of external momenta, $I$ and $J$ with $m$ and $n-m$ elements, respectively ($I\cup J = \{ k_1+z_I\E1, k_2-z_I\E1,k_3, \dots, k_n\}$), and $k_1+z_I\E1\in I$, $k_2-z_I\E1\in J$.

All amplitudes are on shell, however, the tree amplitude in the first term is a forward amplitude and care must be exercised in calculating it. The method we applied in the case of $n=4$ can be generalized to $n\ge 4$ straightforwardly.
Thus, we can reduce the amplitude to three-point amplitudes by a judicious application of the BCFW recursion relations avoiding the singularities. The contribution of the singularities can also be seen to vanish after integration over the loop momentum by a direct calculation, after introducing a momentum regulator.

To define appropriate complex momentum shifts, choose the basis for the polarization vector $\E{n}$,
\be\label{eqEn} \E{n} \in \left\{ \E1 - \frac{\E1\cdot k_n}{k_1\cdot k_n} k_1 \ , \ \E{{n-1}} - \frac{\E{{n-1}}\cdot k_n}{k_{n-1}\cdot k_n} k_{n-1} \right\} ~. \ee
For the choice $\E{n} = \E1 - \frac{\E1\cdot k_n}{k_1\cdot k_n} k_1$, shift
\be k_1 +z_1\E1 \mapsto k_1+z_1\E1 + w \E{n} \ \ , \ \ \ \ k_n \mapsto k_n - w \E{n} ~, \ee
whereas for the choice $\E{n} = \E{{n-1}} - \frac{\E{{n-1}}\cdot k_n}{k_{n-1}\cdot k_n} k_{n-1}$, shift
\be k_{n-1} \mapsto k_{n-1} + w \E{n} \ \ , \ \ \ \ k_n \mapsto k_n - w \E{n} ~. \ee
Notice that there is no need to shift polarization vectors, because $\E{n}\cdot \E1 =0$, and $\E{n}\cdot \E{{n-1}} =0$, respectively.
The contribution from $w\to\infty$ vanishes in both cases and only poles contribute. Thus the $n+2$-point tree amplitude is reduced to lower-point on-shell tree amplitudes. A repetition of this step leads to a reduction to on-shell three-point tree amplitudes.

The final expression (before integrating over the loop momentum) is finite. It should be emphasized that the above reduction process works, and the potential singularity of the forward amplitude is absent, because of the contraction of polarizations of the collinear legs (eq.\ \eqref{eq76a}), without which the forward amplitude would be singular.

\section{Conclusion}
\label{sec:6}

We discussed the applicability of the BCFW recursion relations to the {\em integrand} of loop amplitudes in gauge theories. Working with color-ordered amplitudes, we showed that, with an appropriate choice of basis for the polarization vectors, the contribution from an infinite complex shift can be made to vanish.
Thus, only poles contribute to the loop amplitude. Their residues can be factorized into products of on-shell lower-point loop amplitudes and tree amplitudes. By repeatedly applying the BCFW recursion relations, one thus reduces the loop amplitude to on-shell three-point tree amplitudes.

An obstruction to this reduction procedure is due to one of the poles whose residue is given in terms of a forward amplitude which, in general, contrains singularities. We showed explicitly that the singularities do not contribute to the amplitude, after integrating over the loop momentum. Moreover, by a judicious application of the BCFW recursion relations that we described, potential singularities can be completely avoided. The resulting contribution to the loop amplitude is then written entirely in terms of on-shell three-point tree amplitudes.

It would be interesting to see if our results can be generalized to higher-loop gauge theory amplitudes as well as supergravity. Work in this direction is in progress.

\acknowledgments

We thank U.~al-Binni for discussions. Research supported in part by the Department of Energy under grant DE-FG05-91ER40627.

\end{document}